\documentclass[letterpaper, 10 pt, conference]{ieeeconf}  

\usepackage[version=4]{mhchem}                            
\usepackage{xcolor,colortbl}                              
\usepackage{graphicx}                                     
\usepackage[space]{grffile}	                              
\usepackage{amssymb}                                      
\usepackage{cite} 									      
\usepackage{amssymb}                                      
\usepackage{algorithm}                                    
\usepackage{algpseudocode}                                
\usepackage{mathtools}                                    
\usepackage{blkarray} 									  
\makeatletter
\let\NAT@parse\undefined
\makeatother
\usepackage[colorlinks=true]{hyperref}	

\let\emptyset\varnothing	
\hypersetup{urlcolor=blue,linkcolor=blue,citecolor=blue,colorlinks=true}

\newtheorem{lemma}{Lemma} 
\IEEEoverridecommandlockouts                              
\title{\Large \bf
Realizing Reduced and Sparse Biochemical Reaction Networks from Dynamics}
\author{Maurice Filo and Mustafa Khammash
\thanks{\scriptsize{Maurice Filo ({\ttfamily \scriptsize maurice.filo@bsse.ethz.ch}),  and Mustafa Khammash ( {\ttfamily\scriptsize mustafa.khammash@bsse.ethz.ch}) are with the Department of Biosystems Science and Engineering, ETH Zürich, Switzerland. \newline       
\copyright 2025 IEEE.  Personal use of this material is permitted.  Permission from IEEE must be obtained for all other uses, in any current or future media, including reprinting/republishing this material for advertising or promotional purposes, creating new collective works, for resale or redistribution to servers or lists, or reuse of any copyrighted component of this work in other works.}}
}

\newcommand{\species}[1]{{\bf\ce{#1}}}

\newcommand{\pe}{\theta} 
\newcommand{\p}{\boldsymbol {\theta}} 
\newcommand{\defeq}{\triangleq}
\newcommand{\norm}[2]{\lVert {#1}\rVert_{#2}}
\newcommand{\prox}[2]{\text{prox}_{#1}\left( #2\right)}
\newcommand{\hadamard}{\circ}
\newcommand{\sign}[1]{\text{sign}\left( #1\right)}

\newcommand{\fs}[1]{\mathbb L^2_{#1}[0, T]} 
\DeclarePairedDelimiterX{\inpf}[3]{\langle}{\rangle_{\fs{#3}}}{#1, #2} 
\DeclarePairedDelimiterX{\inpm}[4]{\langle}{\rangle_{\mathbb R^{#3\times#4}}}{#1, #2} 

\newcommand{\opt}[1]{#1^\star}

\begin{document}
\maketitle
\thispagestyle{empty}
\pagestyle{empty}

\begin{abstract} 
We propose a direct optimization framework for learning reduced and sparse chemical reaction networks (CRNs) from time-series trajectory data. In contrast to widely used indirect methods—such as those based on sparse identification of nonlinear dynamics (SINDy)—which infer reaction dynamics by fitting numerically estimated derivatives, our approach fits entire trajectories by solving a dynamically constrained optimization problem. This formulation enables the construction of reduced CRNs that are both low-dimensional and sparse, while preserving key dynamical behaviors of the original system. We develop an accelerated proximal gradient algorithm to efficiently solve the resulting non-convex optimization problem. Through illustrative examples, including a Drosophila circadian oscillator and a glycolytic oscillator, we demonstrate the ability of our method to recover accurate and interpretable reduced-order CRNs. Notably, the direct approach avoids the derivative estimation step and mitigates error accumulation issues inherent in indirect methods, making it a robust alternative for data-driven CRN realizations.
\end{abstract}

\section{Introduction}
Chemical reaction networks (CRNs) offer a unifying framework for modeling a broad class of dynamical processes across disciplines, including systems and synthetic biology \cite{gunawardena2003chemical}, epidemiology \cite{pastor2015epidemic}, and even social and economic systems \cite{dittrich2005reaction}. In the biochemical context, these models describe how concentrations of molecular species evolve over time according to a set of reactions, and they are typically governed by systems of nonlinear ordinary differential equations (ODEs). In many applications, however, the full CRN governing a system's dynamics is unknown, and one seeks instead to infer a reduced interpretable CRN—both in structure and parameters—from experimental or simulated trajectory data.

The growing availability of time-series data has spurred interest in automated methods for inferring reaction networks. Among these, sparse identification approaches \cite{mangan2016inferring, hoffmann2019reactive, bhatt2023sindy, prokop2024biological, jiang2022identification, tuza2019automatic}—many of which build on the principles underlying the sparse identification of nonlinear dynamics (SINDy) algorithm \cite{brunton2016discovering}—have proven particularly effective by leveraging the assumption that a system’s dynamics can be represented as a sparse linear combination of functions from a predefined library. 
Subsequently, several variants of the SINDy algorithm have been developed to improve applicability across different settings \cite{fasel2022ensemble, messenger2021weak}.
These approaches, which we refer to as \textit{indirect} approaches, typically begin by estimating state derivatives from data and then solve a convex optimization problem to identify model parameters. The objective is to minimize the discrepancy between the left- and right-hand sides of the governing dynamical equations—an idea that can be traced back to the work of Bamieh et al. \cite{bamieh2007discovering}.
While computationally efficient, such methods face two key limitations. First, they rely on numerical differentiation, which can introduce significant noise, especially in the presence of measurement error or coarse sampling. Second, they evaluate model quality by matching derivative information rather than trajectories. Crucially, a good fit to the derivatives does not guarantee accurate reproduction of the trajectories themselves—which is usually the primary goal.

To address these challenges, we propose a \textit{direct} approach that realizes CRNs by minimizing the trajectory error itself, subject to the governing ODE constraints. This approach incorporates regularization to promote model sparsity, supports non-negativity constraints to preserve physical interpretability, and enables the inclusion of ``hidden species" to enrich the expressive power of reduced models. While this results in a non-convex optimization problem, we show that it can be effectively addressed using an accelerated proximal gradient algorithm \cite{li2015accelerated} tailored to our setting.
Our contributions are:
\begin{itemize}
    \item We formulate and compare indirect and direct approaches for CRN realization from time-series data, providing a theoretical explanation for the error accumulation issue that may affect indirect methods.
    \item We develop an efficient accelerated proximal gradient descent algorithm for solving the dynamically constrained, sparsity-promoting optimization problem in the direct approach tailored to CRNs. This includes a closed-form expression for the relevant proximal operator, formulated as a straightforward extension to the matrix function setting.
    \item We demonstrate the effectiveness of our method on two benchmark biochemical systems—Drosophila circadian and glycolytic oscillators—showing that it can recover accurate, low-order, and sparse CRNs from high-dimensional, non-mass-action dynamics.
\end{itemize}
Together, these results offer a practical and principled approach to learning interpretable dynamical models from data, with broad implications for reverse engineering biological networks and designing synthetic gene circuits.

\section{Notation}
For a matrix \( \p \in \mathbb{R}^{n \times m} \), we denote its transpose by \( \p^T \), its elementwise \( \ell_1 \)- and \( \ell_2 \)-norms by \( \| \p \|_1 \) and \( \| \p \|_2 \), respectively, and its Frobenius norm by \( \| \p \|_F \). The Hadamard (elementwise) product of two matrices \( \p_1 \) and \( \p_2 \) is denoted \( \p_1 \hadamard \p_2 \). The indicator function \( \mathcal{I}_+(\p) \) returns 0 if all entries of \( \p \) are non-negative and \( +\infty \) otherwise. 
The function \( \sign{\p} \) returns a matrix whose entries are $1$, $-1$, or $0$ depending on whether the corresponding entry in \( \p \) is positive, negative, or zero, respectively.
Let \( I_n \) denote the \( n \times n \) identity matrix. A binary selector matrix \( S \) has entries in \{0,1\}, and its binary complement \( \bar{S} \) is obtained by flipping each entry. The space \( \fs{n} \) denotes square-integrable functions \( [0, T] \to \mathbb{R}^n \).
\( \partial f(x) \) denotes the Jacobian of a function \( f \) evaluated at \( x \).

\section{Problem Setup and Modeling Framework}
This work aims to construct a CRN with a small number of species and reactions, whose dynamics closely replicate a given trajectory \( \{y_T(t)\}_{0 \leq t \leq T} \), where \( y_T(t) \in \mathbb{R}_{\geq 0}^q \).
We consider CRNs consisting of \( n \) species \species{X_1}, \( \cdots \), \species{X_n} and whose dynamics are described by the following system of ODEs:
\begin{equation}
	\dot x_i = f_i(x; \p_i) \defeq \sum_{j=1}^{m} \pe_{ij}\phi_j(x); \quad x_i(0) = x_{i,0}, 
\end{equation}
for $i = 1, \cdots, n$, where each function \( f_i \) on the right-hand side is expressed as a linear combination of \( m \) functions drawn from a library \( \mathcal L_\phi \defeq \{ \phi_1(x), \phi_2(x), \ldots, \phi_m(x) \} \).
The coordinates of \( f_i \) with respect to \( \mathcal{L}_\phi \) are collected in the parameter vector \( \p_i \defeq \begin{bmatrix} \pe_{i1} & \pe_{i2} & \cdots & \pe_{im} \end{bmatrix} \).
The dynamics can be rewritten in a more compact form as:
\begin{equation} \label{Eqn:Main Dynamics}
	\begin{aligned}
		\dot x = f(x;\p) \defeq  \p \Phi(x); \qquad x(0) = x_0, \qquad \\
		\p = \begin{bmatrix} 
			\pe_{11} & \pe_{12} & \cdots &\pe_{1m} \\
			\pe_{21} & \pe_{22} & \cdots & \pe_{2m} \\
			\vdots & \vdots & \ddots & \vdots \\
			\pe_{n1} & \pe_{n2} & \cdots & \pe_{nm}
		\end{bmatrix}, ~~
		\Phi(x) = \begin{bmatrix} 
			\phi_1(x) \\ \phi_2(x) \\ \vdots \\ \phi_m(x)
		\end{bmatrix}.
	\end{aligned}
\end{equation}
We next describe the common approach, in its basic form, for realizing such CRNs from data—a foundation for many existing algorithms \cite{mangan2016inferring, hoffmann2019reactive, bhatt2023sindy, prokop2024biological, jiang2022identification, tuza2019automatic, brunton2016discovering, bamieh2007discovering}—which we shall refer to as the indirect approach. 

{\bf Indirect Approach.} 
This approach realizes CRNs in which the number of species \( n \) matches the dimension \( q \) of the given trajectory \( x_T \defeq y_T \).
It aims to estimate the parameters \( \p \) from the data pair \( \{x_T(t), \dot x_T(t)\}_{0\leq t\leq T} \) while using the smallest possible number of library functions \( \phi_i \) to promote sparsity. 
More precisely, given the data pair, the parameters \( \p \) are estimated by solving the following \(\ell_1\)-regularized constrained optimization problem:
\begin{equation}
	\begin{aligned}
		\opt{\p} = \underset{\p}{\text{argmin}}~ & 
        \frac{1}{2} \int_0^T
        \resizebox{0.575\columnwidth}{!}{$
         \| \dot x_T(t) - \p \Phi(x_T(t)) \|_2^2 dt + \lambda \| \p \|_1 $}\\
		\text{subject to} ~ & S \hadamard \p \geq 0,
	\end{aligned}
\end{equation}
where the time horizon is usually discretized. 
Here, \( S \) is a binary selector matrix that enforces non-negativity constraints on specific parameters to preserve the physical validity of the dynamics.
For example, considering bimolecular CRNs with $n = 3$ species respecting mass-action kinetics, we have 
\begin{equation} \label{Eqn:Library n3}
\begin{aligned}
&\resizebox{0.91\columnwidth}{!}{$
    \Phi(x) = \begin{bmatrix} 1 & x_1 & x_2 &x_3 & x_1^2 & x_1 x_2 & x_1x_3 & x_2^2 & x_2 x_3 & x_3^2 \end{bmatrix}^T$} \\
    &\qquad ~~S= 
    \begin{bmatrix} 
        1 & 0 & 1 & 1 & 0 & 0 & 0 & 1 & 1 & 1 \\
	1 & 1& 0 & 1 & 1& 0 & 1 & 0 & 0 & 1\\
	1 & 1& 1 & 0 & 1& 1 & 0 & 1 & 0& 0\\
    \end{bmatrix}.
\end{aligned}
\end{equation}
It is straightforward to see that this choice of \( S \) enforces nonnegativity on a subset of the parameters in \( \p \), ensuring that the dynamics in \eqref{Eqn:Main Dynamics} remain in the nonnegative orthant.
The cost function consists of two terms: the first is a least-squares error term that quantifies how well $\p$ fits the ODEs, and the second is an \(\ell_1\)-regularization term that promotes sparsity in \( \p \). 
Note that \( \lambda \) is a scalar regularization parameter that controls the trade-off between data fidelity and sparsity in the estimated parameters.
This approach leverages the fact that the right-hand side (RHS) of the dynamical equations is linear in the parameters, rendering the optimization problem convex and allowing for the use of highly efficient numerical solvers. This has enabled the algorithm’s successful application across a range of disciplines.
Next, we describe our approach which we shall call the direct approach. 

{\bf Direct Approach.} 
This approach handles a more general setting in which $n$ is allowed to be larger than $q$, and thus the available data correspond to the trajectory of a subset of the CRN species, denoted by \( y \defeq Cx \), where \( C \in \mathbb{R}^{q \times n} \). 
Given an initial condition \( x_0 \) and a reference trajectory \( \{ y_T(t)\}_{0 \leq t \leq T } \), the direct approach seeks to to estimate \( \p \) by solving the following \(\ell_1\)-regularized, dynamically-constrained optimization problem:
\begin{equation} \label{Eqn:Constrained Optimization Problem}
	\begin{aligned}
		\opt{\p} = \underset{\p}{\text{argmin}} ~ & \frac{1}{2} \int_0^T \| y(t) - y_T(t) \|_2^2 \, dt + \lambda \| \p \|_1 \\
		\text{subject to} ~ &
		\left\{
		\begin{aligned}
			\dot{x} &= \p \Phi(x), \quad x(0) = x_0 \\
			y &= Cx \\
			S &\hadamard \p \geq 0.
		\end{aligned}
		\right.
	\end{aligned}
\end{equation}
This approach generally requires solving a non-convex optimization problem; however, it offers four advantages. 
First, it avoids approximating trajectory derivatives, a process that often introduces significant noise. 
Second, it allows for the inclusion of ``hidden species" in the CRN, which can enrich the system's dynamics and enhance its modeling capacity. 
Third, it does not require the parameters to enter linearly on the RHS of the dynamical equations—this assumption can be readily relaxed and is left for future work. 
Finally, it circumvents a major limitation of the indirect approach where the error dynamically accumulates over time as described next.

{\bf Error Accumulation with the Indirect Approach.}
For simplicity, consider the realization of a linear-affine dynamical system (e.g., a unimolecular CRN) from a given trajectory \( \{y_T(t)\}_{0 \leq t \leq T} \), without enforcing sparsity or positivity constraints. Assume further that \( q = n \). Hence, we have \(\Phi(x) = \begin{bmatrix} x^T & 1 \end{bmatrix}^T, \lambda = 0, S = 0,  C = I_n, x_T \defeq y_T\), and the dynamics are given by:
\begin{equation}
	\begin{aligned}
		\dot{x} = 
		\underbrace{\begin{bmatrix} \pe_{11} & \cdots & \pe_{n1} \\
        \vdots & \ddots & \vdots \\ \pe_{1n} & \cdots & \pe_{nn} \end{bmatrix}}_{A_{\p}} x + 
		\underbrace{\begin{bmatrix} \p_{1, n+1} \\ \vdots \\ \p_{n, n+1} \end{bmatrix}}_{b_{\p}};~ x(0) = x_T(0).
	\end{aligned}
\end{equation}
Define two types of errors:
\begin{equation}
    \begin{aligned}
        e_D(t; \p) &\defeq x(t; \p) - x_T(t), \\
        e_I(t; \p) &\defeq \p \Phi(x_T(t)) - \dot{x}_T(t),
    \end{aligned}
\end{equation}
where \( x(t; \p) \) denotes the solution to \( \dot{x} = A_{\p} x + b_{\p} \) with $x(0) = x_T(0)$ for a given \( \p \).
Observe that \( e_D \) is the error minimized in the direct approach, as it measures the deviation between the simulated trajectory and the target trajectory. 
In contrast, \( e_I \) is minimized in the indirect approach, as it quantifies the mismatch between the parameterized model and the derivatives of the target trajectory.
While both errors are related, we emphasize that minimizing \( e_D \) is more aligned with our ultimate goal: fitting the model-generated trajectory to the given data. 
The indirect error \( e_I \), by comparison, provides only an approximate proxy for this objective.
To understand the relationship between these two errors in the linear-affine setting, consider the dynamics of \( e_D \) 
\begin{equation*}
    \begin{aligned}
        \dot{e}_D &= \dot{x} - \dot{x}_T 
        = A_{\p} x + b_{\p} - \dot{x}_T \\
        &= A_{\p} (x_T + e_D) + b_{\p} - \dot{x}_T 
        = A_{\p} e_D + \p \Phi(x_T) - \dot{x}_T \\
        &= A_{\p} e_D + e_I.
    \end{aligned}
\end{equation*}
This relation yields the following transfer function: $E_D(s) = \left( s I_n - A_{\p} \right)^{-1} E_I(s)$, 
where $s$ is the Laplace variable and $E_D$ and $E_I$ are the Laplace transforms of $e_D$ and $e_I$, respectively. 
This expression reveals that the direct error \( e_D \) evolves according to the same dynamics as the realized system, with the indirect error \( e_I \) acting as the input. As a result, even if the indirect error is minimized to a small value, the direct error—which is the quantity of real interest—may still remain large.
A particularly concerning scenario arises when the system is unstable (i.e., \( A_{\p} \) has eigenvalues in the right-half complex plane or even on the imaginary axis). In such cases, the indirect approach may misleadingly yield a small value of \( e_I \), while the corresponding \( e_D \) grows unbounded over time, severely compromising the quality of the learned model. 
Even when the dynamical system is stable, the closer its eigenvalues are to the imaginary axis, the poorer the quality of the model fit obtained via the indirect approach. To illustrate this, consider a simple birth–death process:
\begin{equation}
    \ce{$\emptyset$ ->[$\mu$] {\textbf{X}_1} ->[$\gamma$] $\emptyset$}, \qquad \dot{x}_1 = -\gamma x_1 + \mu, \quad \mu,\gamma>0.
\end{equation}
Here, we have \( A_{\p} = -\gamma \) and \( b_{\p} = \mu \), leading to a transfer function between the indirect and direct errors given by $\frac{1}{s + \gamma}$.
In particular, the steady-state gain of this transfer function is \( 1/\gamma \), implying that any small residual indirect error \( e_I \)—arising from numerical differentiation, time discretization, or mismatch in the function library—is amplified by a factor of \( 1/\gamma \) in \( e_D \). Thus, for small \( \gamma \), even minor discrepancies in \( e_I \) can lead to significant trajectory mismatches, a fundamental limitation of indirect approaches.

\section{A Proximal Gradient Descent Method for the Direct Approach}
The optimization problem in \eqref{Eqn:Constrained Optimization Problem} can be converted to an unconstrained optimization problem given by
\begin{equation} \label{Eqn:Unconstrained Optimization Problem}
	\begin{aligned}
		&\opt{\p} = \underset{\p}{\text{argmin}}~  J(\p)+  h(\p) \\
		&\text{with} ~
		\left\{
		\begin{aligned}
			J(\p) & = \frac{1}{2} \int_0^T \norm{\left[\mathcal M(\p)\right](t)-  y_T(t)}{2}^2 dt \\
			h(\p) &= \lambda \norm{\p}{1} + \mathcal I_+ (S \hadamard \p) ,
		\end{aligned} \right.
	\end{aligned}
\end{equation}
where $\mathcal M: \mathbb R^{n\times m} \to \fs{n}$ is the parameter-to-output operator defined as 
\begin{equation}
	y = \mathcal M(\p) \Longleftrightarrow 
	\left\{
	\begin{aligned}
		\dot x &= \p \Phi(x); ~~ x(0) = x_0 \\
		y &= Cx.
	\end{aligned}
	\right.
\end{equation}
Recasting the original formulation \eqref{Eqn:Constrained Optimization Problem} as an unconstrained problem yields a cost function with two components: a smooth term \( J(\p) \) and a non-smooth but convex term \( h(\p) \). This structure makes proximal gradient methods \cite{li2015accelerated} particularly well-suited for solving the optimization problem. 
Direct trajectory fitting has a long history in system identification and optimal control. Classical works have used output-error formulations and direct optimization schemes to estimate parameters of ODE models by matching observed trajectories \cite{bock1983recent, betts2010practical}. Our formulation follows a similar principle but focuses on sparsity and CRN structure, which distinguishes it from these earlier methods.
To apply a proximal method, we define the proximal operator for the matrix function \( h: \mathbb{R}^{n \times m} \to \mathbb{R}_{\geq 0} \) as:
\begin{equation} \label{Eqn:Proximal Operator Definition}
	\prox{h,\alpha}{\p} \defeq \underset{\tilde \p}{\text{argmin}} \left\{ \frac{1}{2\alpha} \| \p - \tilde \p \|_F^2 + h(\tilde \p) \right\}, 
\end{equation}
where $\alpha > 0$ is a scalar.
This operator maps a matrix \( \p \) to another matrix \( \hat{\p} = \prox{h,\alpha}{\p}\) that is ``close" to \( \p \) in Frobenius norm while minimizing \( h \). A standard calculation —tailored to the specific structure of our matrix function \( h \)—yields a closed-form expression for the proximal operator, given in the following lemma.
\begin{lemma} \label{Lemma:Proximal Operator}
\textit{
The action of {$\text{prox}_{h,\alpha}: \mathbb R^{n\times m} \to \mathbb R^{n\times m}_{\geq 0}$} defined in \eqref{Eqn:Proximal Operator Definition}, associated with the matrix function \( h \) introduced in \eqref{Eqn:Unconstrained Optimization Problem}, is given by $\hat{\p} = \prox{h,\alpha}{\p}$ with
\begin{equation*} \label{Eqn:Proximal Operator Final Expression}
    \hat{\p} = \max\big\{ S \hadamard \p - \lambda \alpha,\, 0 \big\} 
	+ \sign{ \bar{S} \hadamard \p } \hadamard \max \big\{ | \bar{S} \hadamard \p | - \lambda \alpha,\, 0 \big\},
\end{equation*}
where \( \bar{S} \) is the binary complement of \( S \).}
\end{lemma}
As expected, this operation is essentially a combination of a soft thresholding operation and a projection onto the non-negative orthant. 
The proof can be found in Appendix~B in the supplementary material.

With a closed-form expression for the proximal operator in hand, the proximal gradient method can be applied to solve the unconstrained optimization problem \eqref{Eqn:Unconstrained Optimization Problem} via the following iterative scheme. Given the current estimate \( \p^{(k)} \) of the optimal parameter \( \opt{\p} \), the next iterate \( \p^{(k+1)} \) is computed using the proximal gradient update:
\begin{equation*}
    \p^{(k+1)} = \prox{h,\alpha}{\p^{(k)} - \alpha \nabla J_{\p^{(k)}}}, 
\end{equation*}
where \( \alpha \) is the step size.
All that remains is to calculate the gradient \( \nabla J_{\p^{(k)}} \) evaluated at the current estimate \( \p^{(k)} \). 
A derivation similar to that in \cite{filo2018function, filo2019optimal} is carried out to obtain this gradient and is detailed in Appendix~C in the supplementary material. 
Equipped with the gradient expression, we now present Algorithm~\ref{Algorithm:First Order Algorithm}, which provides a numerical method for realizing CRNs from trajectory data \( \{y_T(t)\}_{0 \leq t \leq T} \) using the direct approach.
\begin{algorithm}[h!]
	\caption{Proximal Gradient Descent Algorithm}
	\label{Algorithm:First Order Algorithm}
	\begin{algorithmic}[1]
		\State Start with an initial guess $\p^{(0)} \in \mathbb R^{n \times m}$ and set $k$ = 0.
		\State Compute the gradient at $\p^{(k)}$, $\nabla J_{\p^{(k)}}$: \newline
		(a) Simulate the dynamics with $\p = \p^{(k)}$:
		\begin{equation*}
		\left\{
		\begin{aligned}
			\dot x^{(k)} &= \p^{(k)} \Phi\big(x^{(k)}\big); ~~x^{(k)}(0) =x_0 \\
			y^{(k)}&= C x^{(k)}.
		\end{aligned}
		\right.
		\end{equation*}
		(b) Compute the time-varying vector $b(t)$ and matrix $A(t)$:
		\begin{align*}
			b^{(k)}(t) &= \Phi\big[ x^{(k)}(t)\big], & 
			A^{(k)} (t)&= \p^{(k)} \partial \Phi\big(x^{(k)}\big).
		\end{align*}
		(c) Solve for $\lambda^{(k)}(t)$, with $\lambda^{(k)}(T) = 0$:
		\begin{align*}
			\dot{\lambda}^{(k)} &=- \left[A^{(k)}\right]^T \lambda^{(k)}- 
			C^T\left(y^{(k)}- y_T\right).
		\end{align*}
		(d) Compute the gradient:
		\begin{equation*}
			\nabla J_{\p^{(k)}} = \int_0^T \lambda^{(k)}(t) \left[b^{(k)}(t)\right]^T dt.
		\end{equation*}
		\State Proximal gradient update with step $\alpha$:
		\begin{equation*} 
			\begin{aligned}
				\tilde \p^{(k)} &= \p^{(k)} - \alpha \nabla J_{\p^{(k)}}\\
				\p^{(k+1)} &= \text{max} \Big\{S \hadamard \tilde \p^{(k)} - \lambda \alpha, 0\Big\} \\
				&+  \sign{\bar S \hadamard \tilde \p^{(k)}} \hadamard  \text{max} \Big\{|\bar S \hadamard\tilde \p^{(k)}| - \lambda \alpha, 0\Big\},
			\end{aligned}
		\end{equation*}
		\State  Set $k= k + 1$ and go back to step 2. Repeat until convergence.
	\end{algorithmic}
\end{algorithm}
Algorithm~\ref{Algorithm:First Order Algorithm} is presented in its basic first-order form, but several enhancements can improve its performance. The step size \( \alpha \) can be made adaptive, for example via the Armijo rule \cite{armijo1966minimization}; second-order information can be incorporated for Newton-type updates; and acceleration can be achieved by introducing momentum. In our numerical examples, we use an accelerated proximal gradient method \cite{li2015accelerated}, where in Step 3, the parameter \( \p^{(k)} \) is replaced by a momentum-corrected estimate \( \p^{(k)}_m \), defined as:
\begin{equation}
    \begin{aligned}
        \p^{(k)}_m &= \p^{(k)} + \frac{\zeta_{k-1} - 1}{\zeta_k} \left( \p^{(k)} - \p^{(k-1)} \right), \\
        \zeta_{k+1} &= \frac{1 + \sqrt{1 + 4\zeta_k^2}}{2},
    \end{aligned}
\end{equation}
with initialization \( \zeta_0 = 0 \) and \( \p^{(0)}_m = \p^{(0)} \).

\section{Numerical Examples}
This section presents two numerical examples where trajectories from high-dimensional bioCRNs—potentially governed by non–mass-action kinetics—are used to construct reduced models that follow mass-action kinetics. This presents a nontrivial challenge, as the algorithm must approximate complex dynamics using a more constrained and  more easily interpretable reaction structure. The first example is a Drosophila circadian oscillator model from \cite{leloup1999chaos}, and the second is a glycolytic oscillator model from \cite{daniels2015efficient}.

\subsection{Drosophila Circadian Oscillator}
We begin with the Drosophila circadian oscillator, governed by the interactions of the Period (PER) and Timeless (TIM) proteins through phosphorylation and negative feedback from a nuclear PER-TIM complex on the transcription of the \textit{per} and \textit{tim} genes. Based on experimental evidence, \cite{leloup1999chaos} proposed a model capturing these dynamics and showing the emergence of chaos and birhythmicity. This model has since been widely used to study nonlinear behavior in circadian systems \cite{leloup1998model, tsumoto2006bifurcations, bamieh2007discovering}.
In our study, we adopt both the model structure and parameter values from \cite[Eq.~(1)]{leloup1999chaos}, which involves 10 species and combines mass-action kinetics with Hill-type nonlinearities. While \cite{bamieh2007discovering} focused on identifying a dense, lower-order model that captures the original dynamics, we show that our method can produce models that are both low-order and sparse, yet still accurate.

\begin{figure}[ht!]
\includegraphics[width=0.48\textwidth]{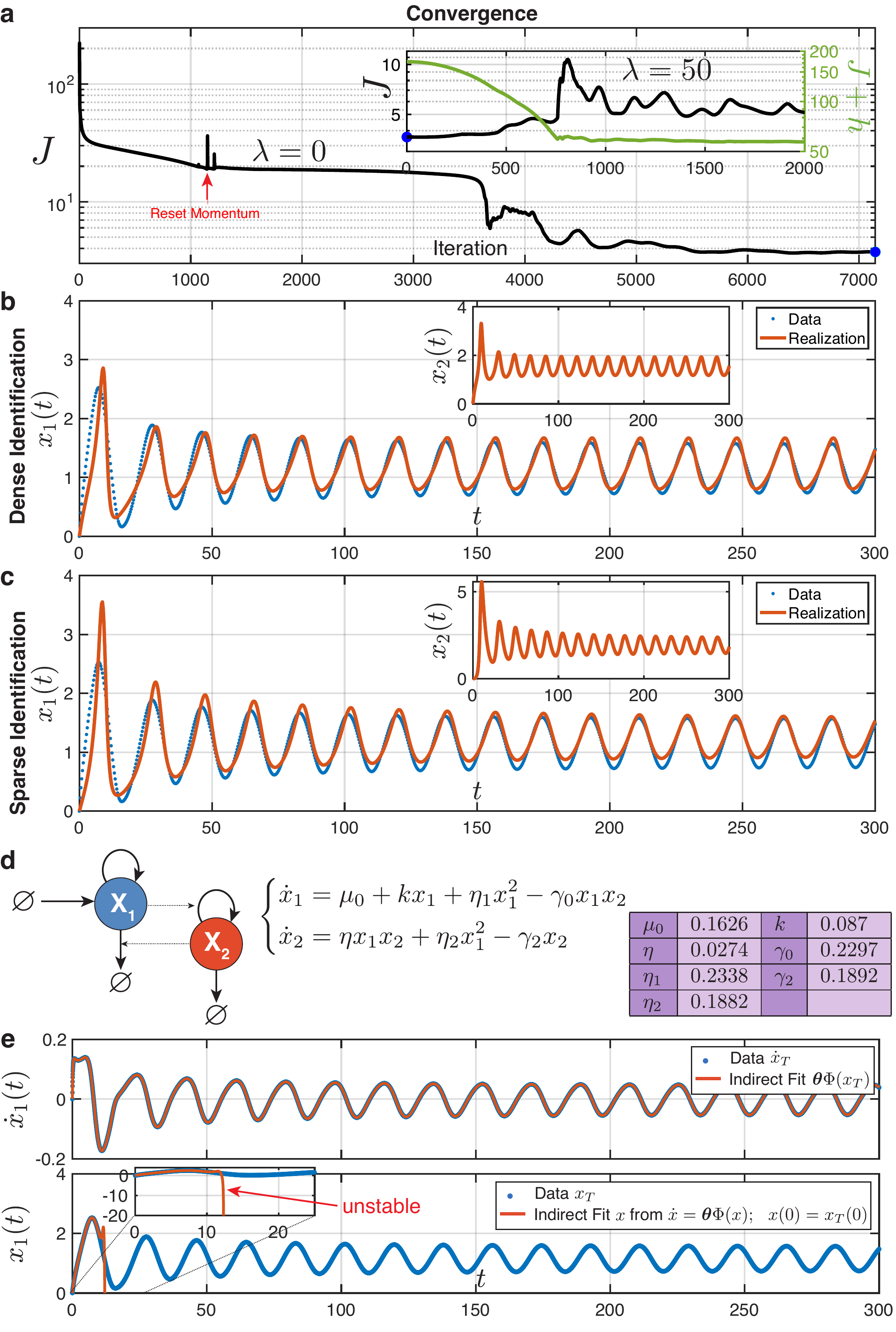}
\vspace{-0.4cm}
\caption{Reduced bioCRN realization of a drosophila circadian oscillator trajectory. (a) Convergence plots: the main plot shows the algorithm run with \( \lambda = 0 \), while the inset shows the convergence with \( \lambda = 50 \), initialized from the final iterate of the unregularized run. (b) and (c) Comparison between the original full model and the reduced models: (b) dense, and (c) sparse. (d) Topology and dynamical equations of the sparsely identified bioCRN. (e) Full model fitting ($n=10$) with the indirect method.}
\vspace{-0.6cm}
\label{Fig:PERTIM}
\end{figure}

We assume that the trajectory of the first species, \( x_1 \), from the original model \cite[Eq.~(13)]{bamieh2007discovering} is available, i.e., the measured output is \( y_T = x_1 \). The objective is to construct a reduced bimolecular bioCRN with \( n = 2 \) species using the function library and binary selector matrix: 
\begin{equation}
\begin{aligned}
    \Phi(x) &= \begin{bmatrix} 1 & x_1 & x_2 & x_1^2 & x_1 x_2 & x_2^2 \end{bmatrix}^T \\
    S &= \begin{bmatrix}
        1 & 0 & 1 & 0 & 0 & 1\\
        1 & 1 & 0 & 1 & 0 & 0
    \end{bmatrix},
\end{aligned}
\end{equation}
such that the trajectory of the first species closely matches the observed trajectory. The second species is a hidden variable that is introduced to enhance the modeling capacity of the reduced network.
We begin by setting \( \lambda = 0 \), ignoring sparsity and thus identifying a dense model. The accelerated proximal gradient algorithm is run with a fixed step size \( \alpha = 10^{-7} \) until convergence, stopping at \( k_{\max} = 7100 \) iterations. The non-regularized cost \( J \) is plotted in Fig.~\ref{Fig:PERTIM}(a). The momentum is reset once—marked by a red arrow—when a plateau in the cost function is detected.
The resulting dense parameter matrix \( \p \) is provided in Appendix~A in the supplementary material.
Fig.~\ref{Fig:PERTIM}(b) shows the corresponding simulation results, where blue denotes the original full-model trajectory and red shows the response of the reduced model. The reduced system achieves a strong match despite the significant drop in dimensionality.

To promote sparsity, we re-run the algorithm using \( \lambda = 50 \), initializing with the previously obtained dense matrix and running for 2,000 iterations. The convergence of the total cost \( J + h \) and the fidelity term \( J \) are shown in the inset of Fig.~\ref{Fig:PERTIM}(a). As expected, enforcing sparsity slightly increases the fit error \( J \), reflecting the inherent trade-off between simplicity and accuracy.
The resulting sparse bioCRN, along with its governing equations, is shown in Fig.~\ref{Fig:PERTIM}(d), and the simulation results are plotted in Fig.~\ref{Fig:PERTIM}(c). Although slightly less accurate than the dense model, the sparse model still captures the dynamics well—especially given the dramatic reduction in complexity (2 species and 7 reactions, versus 10 species and 44 reactions in the original system).
Finally, we note that the regularization parameter \( \lambda \) serves as a tuning knob, mediating the balance between accuracy and sparsity.

To illustrate the limitations of the indirect approach on oscillatory dynamics, we applied it to fit the full 10-species circadian oscillator model using bimolecular mass-action kinetics but with no regards to sparsity. As shown in Fig.~\ref{Fig:PERTIM}(e), only the first species is displayed. While the top panel shows an accurate fit of the time derivative ($\dot x_T$ and $\p\phi(x_T)$), the trajectory (bottom panel) matches the dynamics only briefly before diverging and becoming unstable—an outcome attributed to the error accumulation problem discussed earlier.

\subsection{Glycolytic Oscillator}
Glycolytic oscillations are periodic fluctuations in the concentrations of metabolic intermediates, first observed in 1957 \cite{duysens1957fluorescence}. More recently, they have become a standard benchmark for model prediction, system identification, and automated inference \cite{brunton2016discovering, daniels2015automated, daniels2015efficient}. In this study, we adopt the model structure and parameters from \cite[Eq.~(19)]{daniels2015efficient}, which involves seven species and combines mass-action kinetics with Hill-type nonlinearities.
This model was also examined in \cite[SI Appendix B]{brunton2016discovering} to highlight a limitation of their sparse identification framework which follows an indirect approach. The authors reported that, even when using a mass-action bioCRN with the correct number of species (i.e. same model order), the identified model produced derivatives that closely matched the measured ones. However, they demonstrate that the resulting dynamics deviated from the true system beyond the initial phase of the simulation. We attribute this discrepancy to the error accumulation problem. In contrast, we show that our direct approach mitigates this issue: even reduced-order bioCRNs with fewer species can accurately capture the dynamics, including for species governed by Hill-type kinetics—an aspect that posed a significant challenge in the approach of \cite{brunton2016discovering}.

In this example, we assume access to the trajectories of the first two species, \( x_1 \) and \( x_2 \), from the original model (see \cite[SI Appendix B]{brunton2016discovering}), i.e., the measured output is \( y_T = \begin{bmatrix} x_1 & x_2 \end{bmatrix}^T \). The objective is to construct a reduced bimolecular bioCRN with \( n = 3 \) species—using the function library and binary selector matrix in \eqref{Eqn:Library n3}—such that the trajectories of \( x_1 \) and \( x_2 \) closely match the data. The third species serves as a hidden variable.
As before, we first set \( \lambda = 0 \). The accelerated proximal gradient algorithm is run with \( \alpha = 10^{-5} \) for \( 14{,}000 \) iterations. The convergence of the cost \( J \) is shown in Fig.~\ref{Fig:Glyco}(a), with four momentum resets. 
The final \( \p \) is listed in Appendix~A, and the simulation results in Fig.~\ref{Fig:Glyco}(b) show excellent agreement with the original trajectories, despite the reduced model order.
We then use the identified dense \( \p \) as an initial guess and rerun the algorithm with \( \lambda = 0.05 \) to promote sparsity, stopping after 10,000 iterations. The convergence of both the total cost \( J + h \) and the fidelity term \( J \) is shown in the inset of Fig.~\ref{Fig:Glyco}(a). The resulting sparse bioCRN and its governing equations are shown in Fig.~\ref{Fig:Glyco}(d), with the corresponding simulation results in Fig.~\ref{Fig:Glyco}(c). While the sparse model is slightly less accurate, it still captures the dynamics well—especially considering the reduced complexity (3 species and 15 reactions vs. 7 species and 26 reactions in the original system).
\begin{figure}[ht!]
\includegraphics[width=0.48\textwidth]{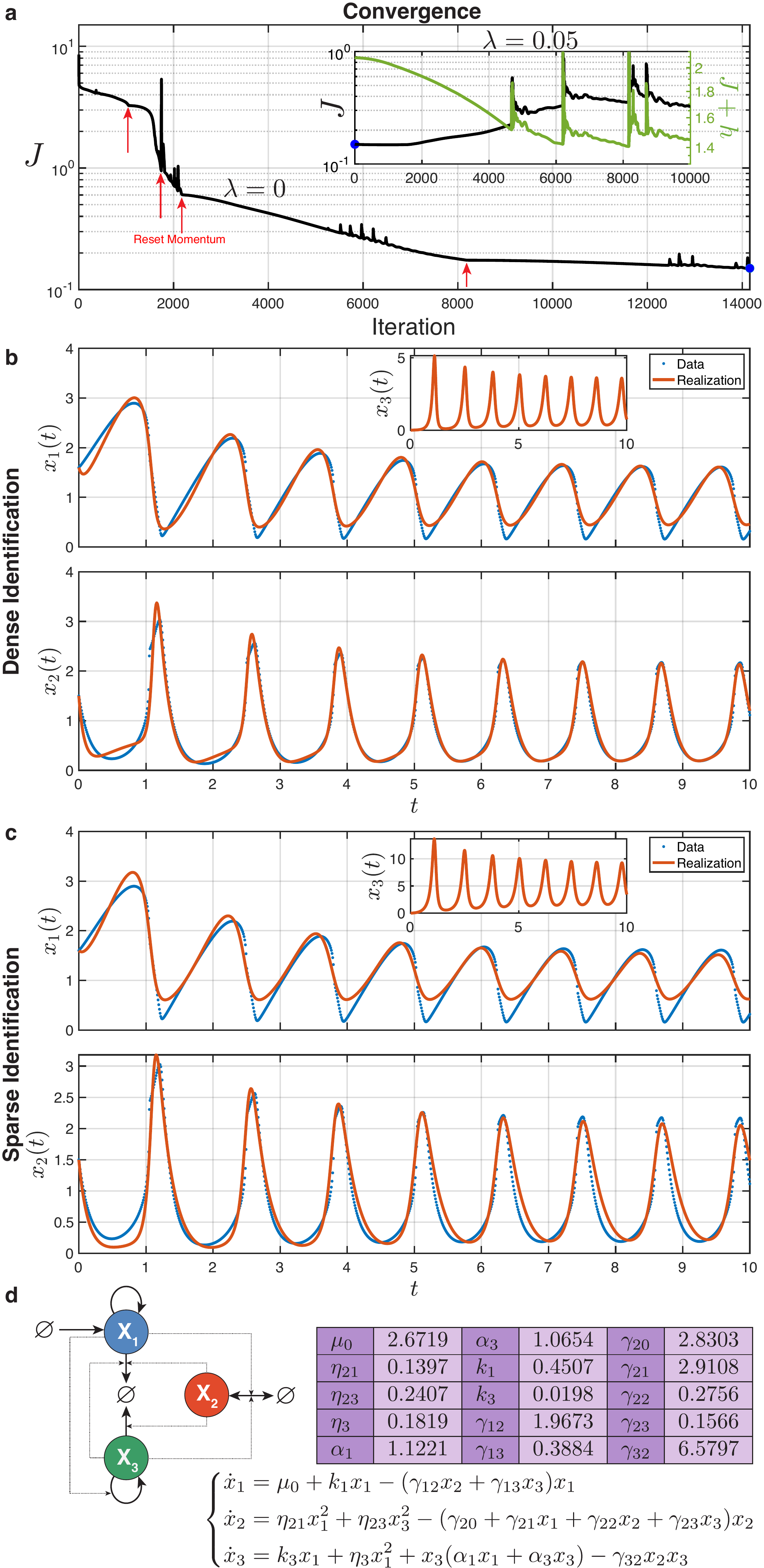}
\vspace{-0.3cm}
\caption{Reduced bioCRN realization of a glycolytic oscillator trajectory.}
\vspace{-0.65cm}
\label{Fig:Glyco}
\end{figure}

\section{Discussion} 
This paper introduced a direct optimization framework for realizing reduced and sparse CRNs from trajectory data. The goal is not to dismiss indirect (derivative-based) methods, which remain powerful and computationally efficient, but to highlight a key limitation: when the underlying dynamics are near instability, small errors due to numerical computations or mismatch in the library functions can accumulate, leading to poor trajectory fits despite accurate ODE residuals. In such cases, the direct method we propose is more robust, as it fits the trajectories directly and avoids derivative approximation altogether. The trade-off is that it requires solving a non-convex optimization problem, in contrast to the convex formulations typical of indirect approaches.
The proximal gradient algorithm we presented here is a basic first-order scheme, but it opens the door to many improvements. These include adaptive time stepping, second-order methods using Hessians, adding biological constraints, and drawing from recent enhanced variants of SINDy \cite{fasel2022ensemble, messenger2021weak, hoffmann2019reactive}. Furthemore, our method can be easily extended to handle multiple trajectories arising from different initial conditions or inputs. Incorporating such features could further improve accuracy, interpretability, and robustness.
Overall, the proposed direct approach complements existing methods and is especially well-suited for near-unstable systems, where indirect methods may struggle. 

\bibliographystyle{IEEEtran}

\end{document}